\documentclass[aps,prl,preprint,groupedaddress,showkeys,showpacs]{revtex4}
\usepackage{amsmath}
\usepackage{graphicx}
\begin{document}


\title{On the Pulsar Emission Mechanism }


\author{Paolo Cea$^{1,2,}$}
\email[]{Paolo.Cea@ba.infn.it}
\affiliation{$^1$Physics Department, Univ. of Bari, I-70126 Bari, Italy \\
$^2$INFN - Sezione di Bari, I-70126 Bari, Italy}

%
\begin{abstract}
We discuss a general mechanism which allows to explain naturally
both radio and high energy emission by pulsars. We also discuss
the plasma distribution in the region surrounding the pulsar, the
pulsar wind and the formation of jet along the magnetic axis. We
suggest a plausible mechanism to generate pulsar proper motion
velocities.
\end{abstract}

\pacs{97.60.Gb, 95.30.-k,}
\keywords{ Pulsar, Emission Mechanism, Magnetic Field}

\maketitle

\begin{figure}
\includegraphics[width=0.90\textwidth,clip]{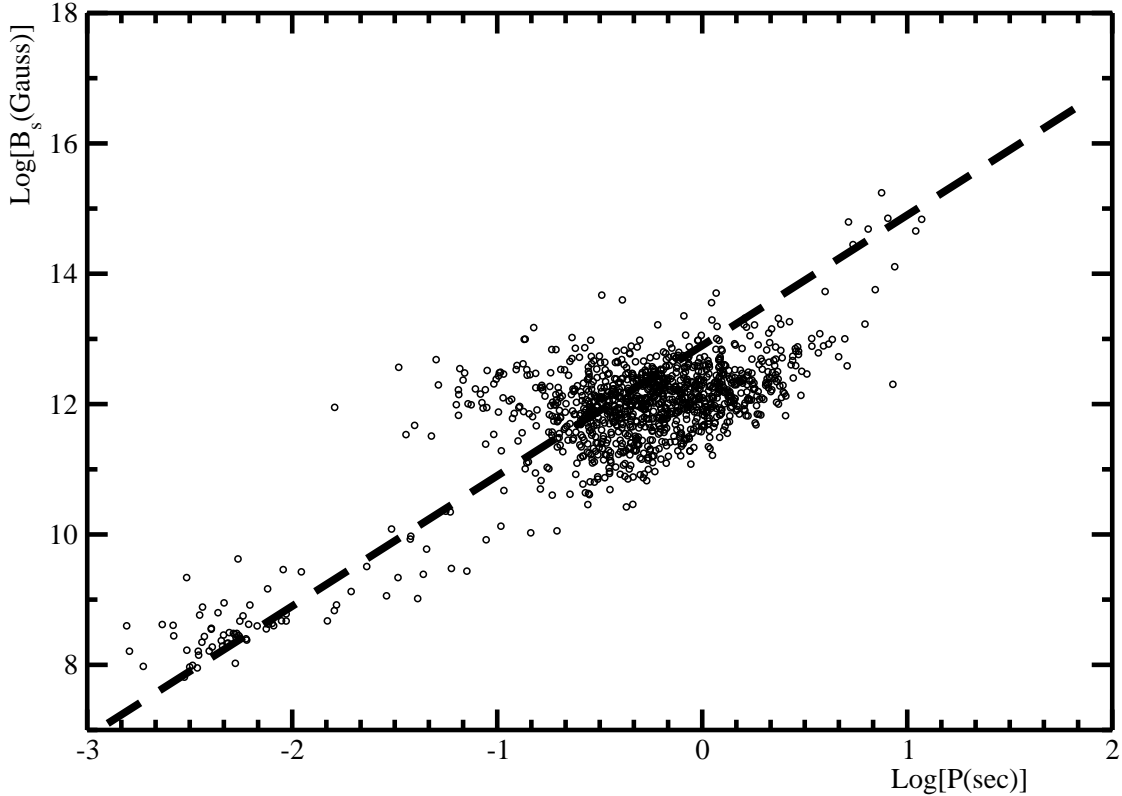}
\caption{\label{fig:01}
Inferred magnetic field $B_S$  plotted versus stellar period for
1194 pulsars taken from the ATNF Pulsar Catalog~\cite{ATNF}.
Dashed line corresponds to Eq.~(\ref{magn-period}) with $B_1 \,
\simeq \, 1.3 \, 10^{13} \; Gauss$.}
\end{figure}

%
%
%
One of the most fundamental discoveries in astrophysics was made
with the  discovery of pulsars in 1967~\cite{hewish:1968}. Pulsars
have been soon identified with neutron stars, first predicted
theoretically by W.~Baade and
F.~Zwicky~\cite{baade:1934,manchester:1977}. Despite theoretical
efforts over more than 30 years since the discovery of the first
radio pulsar, the pulsar emission mechanism is still a challenge
to astrophysics. Even thought it is generally accepted that
pulsars are rapidly rotating neutron stars endowed with a strong
magnetic field~\cite{pacini:1968,gold:1968},  the exact mechanism
by which a pulsar radiates the energy observed as radio pulses is
still a subject of vigorous
debate~\cite{michel:1982,michel:1991,meszaros:1992}.
Notwithstanding  the accepted standard model based on the picture
of a rotating magnetic dipole has been developed since long
time~\cite{goldreich:1969,sturrock:1971}. \\
Recently, we have proposed~\cite{cea:2003} a new class of compact
stars, named P-stars, which is challenging the two pillars of
modern astrophysics, namely neutron stars and black holes. Indeed,
in our previous paper~\cite{cea:2003} we showed that, if we assume
that pulsars are P-stars, then we may completely solve the
supernova explosion problem. Moreover, we found that cooling
curves of P-stars compare rather well with available observational
data. We are, however, aware that such a dramatic change in the
standard paradigm of relativistic astrophysics needs a careful
comparison with the huge amount of observations collected so far
for pulsars and black hole candidates. As a first step in this
direction, in Ref.~\cite{cea:2004a} we showed that P-stars once
formed are absolutely stable, for they  cannot decay into neutron
or strange stars. Moreover, we convincingly argued that the
nearest isolated compact stars {\it {RXJ1856.5-3754}} and {\it
{RXJ0720.4-3125}} could be interpreted as P-stars with $M \,
\simeq 0.8 \, M_{\bigodot}$ and
$R \, \simeq 5 \, Km$. \\
In a forthcoming paper~\cite{cea:2004b} we will address the
problem of generation  of the magnetic field and we will discuss
the glitch mechanism in P-stars. In particular, we will argue that
 P-stars with canonical mass $ M \, \simeq \, 1.4 \, M_{\bigodot}$
are allowed to generate dipolar surface magnetic fields up to $B_S
\, \simeq \, 10^{17} \; Gauss$. As we will discuss in
Ref.~\cite{cea:2004b}, the generation of the dipolar magnetic
field is enforced by the formation of a dense inner core composed
mainly by down quarks in the $n=1$ Landau levels. In general the
formation of the inner core denser than the outer core is
contrasted by the centrifugal force produced  by stellar rotation.
This leads us to suppose that the surface magnetic field strength
is proportional to the square of the star spin period:
\begin{equation}
\label{magn-period}
B_S\; \; \simeq \; \; B_1 \; \left (\frac{P}{1 \, sec } \right )^2
\; \; \; \; ,
\end{equation}
where $B_1$ is the surface magnetic field for pulsars with period
$P \, = \, 1 \; sec$. Remarkably, assuming $B_1 \; \simeq \;
 1.3 \; \;  10^{13}  \; \;  Gauss$, we find the Eq.~(\ref{magn-period})
accounts rather well for the inferred magnetic field of pulsars
ranging from millisecond pulsars up to anomalous $X$-ray pulsars
and soft-gamma repeaters.
Indeed, in Fig.~\ref{fig:01} we display the surface  magnetic
field strength $B_S$ (for instance, see
Ref.~\cite{manchester:1977}):
\begin{equation}
\label{magn-surf}
B_S\; \; \simeq \; \; 3.1 \; 10^{19} \; \; \sqrt{P \;\dot{P} }  \;
Gauss \; \; ,
\end{equation}
versus the period. We see that our Eq.~(\ref{magn-period}) appears
to descrive the inferred surface magnetic field strength of most
pulsars fairly well, although there is some scatter in the data.
\\
A straightforward consequence of Eq.~(\ref{magn-period}) is that
the dipolar magnetic field is time dependent. In fact, it is easy
to find:
\begin{equation}
\label{mag-time}
 B_S(t) \; \simeq \;  B_0 \; \; \left ( 1 \; + \; 2 \; \frac{\dot{P}}{P} \; t \;  \right )
 \; \;  \; \; ,
\end{equation}
where $B_0$ indicates the magnetic field at the initial
observation time. Note that Eq.~(\ref{mag-time}) implies that the
magnetic field varies on a time scale given by the characteristic
age:
\begin{equation}
\label{char-age}
 \tau_c \; = \; \frac{P}{2 \, \dot{P}} \; \;.
\end{equation}
A remarkable consequence of Eq.~(\ref{mag-time}) is that the
effective braking index $n$ is time dependent. In particular, the
braking index decreases with time such that:
\begin{equation}
\label{braking}
 -1 \; \lesssim \; n \; \lesssim \; 3  \; \;,
\end{equation}
the time scale variation being of order of $\tau_c/2$. However, it
turns out that~\cite{cea:2004b} the monotonic derive of the
braking index is contrasted by the glitch activity. Indeed, in our
theory the glitches originate from dissipative effects in the
inner core of the star leading to a decrease of the strength of
the dipolar magnetic field, but to an increase of the magnetic
torque. Moreover, we find that the time variation of the dipolar
magnetic field could account for pulsar timing noise.
In the present paper we would like to discuss a fair general
mechanism which allows to explain naturally  pulsar radio emission
as well as  high energy emission. Our analysis is based on the
remarkable dependence of the dipolar magnetic field on the spin
period, Eq.~(\ref{magn-period}), which seems to be supported by
observational data. Even though such a dependence is natural
within our P-star theory, one could follow a more pragmatic
approach and merely assume the validity of Eq.~(\ref{magn-period})
as a reasonable description of pulsar data. \\
In polar coordinate, the pulsar  dipolar magnetic field
$\vec{B}(\vec{r})$  for $r  >  R$, $R$ being the radius of the
star, is given by:
\begin{equation}
\label{mag-polar}
\begin{split}
 B_r & = \; - \frac{2 \; m \; \cos \theta}{r^3}  \; \;  ,  \\
 B_\theta & = \; - \frac{ m \; \sin \theta}{r^3}  \; \;  ,  \\
B _\varphi & =  \; 0 \; ,
\end{split}
\end{equation}
where:
\begin{equation}
\label{mag-mom}
 m \; = \; B_S(t) \; R^3  \;
\end{equation}
is the magnetic moment. Note that, according to previous
discussion we are assuming that the surface magnetic field
strength is time dependent. Thus, from Maxwell equations
\begin{equation}
\label{maxwell}
 \vec{\nabla} \; \times \; \vec{E} \; = \;
 - \; \frac{\partial \; \vec{B}}{\partial \;t}  \; ,
\end{equation}
where  natural units $c = 1$ are used, it follows:
\begin{equation}
\label{electric}
E_\varphi  =  + \frac{ \dot{m}  \; \sin \theta}{r^2}  \; \; \; ,
\; \; \; r \; > \; R \; \; \; .
\end{equation}
Note that, according to Eq.~(\ref{mag-time}) we have:
\begin{equation}
\label{mom-time}
 \dot{m}  \; \simeq \; \; 2 \; B_0 \; R^3  \;  \frac{\dot{P}}{P} \; \; \;
 ,
\end{equation}
so that:
\begin{equation}
\label{electric-full}
E_\varphi  \; \simeq \; \; 2 \; B_0 \; \sin \theta \; \;
\frac{\dot{P}}{P} \; \; \frac{R^3}{r^2} \;   \; \; , \; \; \; r \;
> \; R \; \; \; .
\end{equation}
It is worthwhile  to note that the induced azimuthal electric
field Eq.~(\ref{electric-full}) depends on the stellar period and
period derivative. As we discuss below, it is this electric
field which accounts for both radio and high energy emission. \\
In the following we work in the co-rotating frame of the star. We
also assume that the magnetosphere contains a plasma whose charge
number density is approximately the Goldreich-Julian charge
density~\cite{michel:1991,meszaros:1992}. These charges are
accelerated by the induced azimuthal electric field $E_\varphi$
and thereby acquire an azimuthal velocity $v_\varphi$ which is
directed along the electric field for positive charges and in the
other direction for negative charges. Note that we do not need to
separate positive charges from the negative ones, in other words
we do not feel the current closure problem~\cite{michel:1991}.
Charged particles moving in the magnetic field $\vec{B}(\vec{r})$,
Eq.~(\ref{mag-polar}),  must emit electromagnetic waves, namely
cyclotron radiation for non relativistic charges or synchrotron
radiation for relativistic charges~\cite{wallace:1977}. Obviously,
radiation from electrons is far more important than from protons.
So that in the following we shall consider electrons. For
convenience, we also assume that electrons have  positive charge
$+e$. It turns out that electron cyclotron emission accounts for
radio emission, while the synchrotron radiation is responsible for
the high energy emission. \\
However, before addressing the problem of the emission spectra, it
is worthwhile to discuss the distribution of the plasma in the
region surrounding the pulsar (the magnetosphere). As we said
before, charges are accelerated by the electric field $E_\varphi$,
so that they are subject to the drift Lorentz force $\vec{F} \, =
\, \vec{v}_\varphi \, \times \, \vec{B}(\vec{r})$, whose radial
component is:
\begin{equation}
\label{F_r}
 F_r  = -  v_\varphi  B_\theta  =   +  v_\varphi \;
  \frac{ m \; \sin \theta}{r^3}   \simeq  + \; v_\varphi B_0
  \sin \theta  \left ( \frac{ R}{r} \right )^3  ,
\end{equation}
while the $\theta$ component is:
\begin{equation}
\label{F_theta}
F_\theta  =  +  v_\varphi  B_r = - v_\varphi
  \frac{ m  \cos \theta}{r^3} \simeq -  v_\varphi
 B_0  \cos \theta  \left ( \frac{
R}{r} \right )^3  .
\end{equation}
The radial component $F_r$ pushes both positive and negative
charges radially outward. Then, at large distances form the star
the plasma must flow radially outward giving rise to the pulsar
wind. On the other hand, $F_\theta$ leads to an asymmetric charge
distribution in the upper hemisphere ($\cos \theta > 0$) with
respect to the lower hemisphere ($\cos \theta < 0$). Indeed,
$F_\theta$ is centripetal in the upper hemisphere and centrifugal
in the lower hemisphere. As a consequence, in the lower hemisphere
charges are pushed towards the magnetic equatorial plane $\cos
\theta = 0$. On the other hand, in the upper hemisphere the
centripetal force gives rise to a rather narrow jet along the
magnetic axis. Therefore, the emerging plasma distribution is a
follows: a rather broad structure in the lower hemisphere, a
narrow jet in the upper hemisphere, and an accumulation of plasma
near the magnetic equatorial plane. Moreover, the drift force
causes a continuous  injection of charges from the lower
hemisphere into the upper hemisphere. This, in turns, results in
formation of plasma waves of large intensities propagating along
the magnetic equatorial plane. It is remarkably that our
qualitative discussion of plasma distribution around the pulsar
turns out to compare rather well with recent observations of Crab
and Vela
pulsars~\cite{pavlov:2001,helfand:2001,hester:2002,gaensler:2002,pavlov:2003},
after identifying the observed symmetry axis with the magnetic
axis, and not with the rotation axis as usually assumed. Finally,
it is worthwhile to point out that a fraction of the plasma
injected into the upper hemisphere are eventually accelerated into
the narrow jet. This process results into an acceleration of the
pulsar giving a proper velocity directed along the magnetic axis
and pointing in the direction opposite to the narrow jet. \\
Any further discussion of these points goes beyond the aim of the
present paper. Here we shall focus on the emission processes
capable of producing radiation at both radio and high energy
frequencies. In particular, we do not attempt any precise
comparison with available data, but we limit to estimate the
relevant luminosities for typical pulsars. \\
In the following we assume as typical pulsar parameters:
\begin{equation}
\label{typical}
\begin{split}
 P & \; \simeq  \;  1 \; sec \; \; , \; \; \dot{P} \; \simeq  \; 10^{-14} \;  , \;
 B_0  \; \simeq \; 10^{12} \; Gauss \\
 R & \; \simeq \;   10^6 \; cm \; \; , \; \; I \; \simeq \;   10^{45} \; gr \; cm^2
  \;  .
\end{split}
\end{equation}
From Eq.~(\ref{electric-full}) it follows that electrons acquire
non relativistic azimuthal velocities farther out of the star,
while they are relativistic near the star. So that non
relativistic electrons moving in the magnetic field will emit
cyclotron radiation, while  relativistic electrons will emit
synchrotron radiation. In both cases the radiated power is
supplied by the azimuthal electric field $E_\varphi$. Let us,
first, consider the cyclotron emission. As it is well known, most
of the radiation is emitted almost at the frequency of rotation:
\begin{equation}
\label{cyclo}
 \nu(r)  =  \frac{e  B(r)}{2 \pi m_e}  \simeq
 \frac{e B_0  }{ 2 \pi m_e}
   \left ( \frac{ R}{r} \right )^3  ,
\end{equation}
where $r$ is the radial distance from the star. From
Eq.~(\ref{cyclo}) it follows:
\begin{equation}
\label{r-nu}
 r  \simeq  R \left ( \frac{e  B_0}{ 2 \pi m_e}  \right )^{1/3}
 \frac{1}{\nu^{1/3}}  \; .
\end{equation}
To estimate the total power emitted we need to evaluate the power
supplied by the azimuthal electric field. In the infinitesimal
volume $dV=r^2 \sin \theta dr d\theta d\varphi$ the power supplied
by the induced electric field $E_\varphi$ is:
\begin{equation}
\label{inf-pow}
 d \dot{W}_{E_\varphi} \simeq  2 n_e e B_0 v_\varphi   R^3  \;  \frac{\dot{P}}{P}
 \sin^2 \theta dr d\theta d\varphi  \; ,
\end{equation}
where $n_e$ is  the electron number density. Now, from
Eq.~(\ref{r-nu}) we get:
\begin{equation}
\label{dr-dnu}
 dr  \simeq \frac{R}{3} \left ( \frac{e  B_0}{ 2 \pi m_e}  \right )^{1/3}
 \frac{1}{\nu^{4/3}} \; d\nu \; .
\end{equation}
So that, integrating $d \dot{W}_{E_\varphi}$ over $\theta$ and
$\varphi$, and using Eq.~(\ref{dr-dnu}), we obtain the radiating
spectral power:
\begin{equation}
\label{spect-pow}
 F(\nu) \simeq  \frac{2 \pi^2}{3}  n_e e B_0 v_\varphi   R^4  \;
 \frac{\dot{P}}{P} \left ( \frac{e  B_0}{ 2 \pi m_e}  \right )^{1/3}
 \frac{1}{\nu^{4/3}} \; .
\end{equation}
Few comments are in order. Firstly, in our model higher
frequencies are generated near the star while lower frequencies
further out. Indeed, our Eq.~(\ref{r-nu}) gives a radius to
frequency mapping which seems to be consistent with observations.
The authors of Ref.~\cite{kijak:1997} (for a recent review see
Ref.~\cite{graham-smith:2003} and references therein) argued that
the emission altitude $r_h$ depends on pulsar period, period
derivative and frequency according to:
\begin{equation}
\label{rhfm}
r_h  =  \left (55 \pm 5 \right ) R \; \tau_6^{-0.07 \pm 0.3} \;
P^{0.33 \pm 0.5} \; \nu_{GHz}^{- 0.21 \pm 0.07} \; ,
\end{equation}
where $\tau_6$ is the characteristic age in units of $10^6$ years
and $\nu_{GHz}$ is the frequency in units of $10^9 \, Hz$. On the
other hand, using Eqs.~(\ref{magn-surf}),~(\ref{r-nu}) we get:
\begin{equation}
\label{rfm}
 r \; \simeq  \; \left (2.1 \; 10^3 \right ) R \;  \; \dot{P}_{14}^{1/6}
 \; P^{1/6}
 \; \nu_{GHz}^{- 1/3 } \; ,
\end{equation}
where $\dot{P}_{14}$ is the period derivative in units of
$10^{-14}$. The altitude $r_h$ is related to radial distance $r$
by:
\begin{equation}
\label{r-r_h}
 r_h  \; \simeq \; r \; \cos \theta \; \; .
\end{equation}
We see that, if $\cos \theta \lesssim 0.1$ our Eq.~(\ref{rfm}) is
in reasonable agreement with the semi-empirical relation
Eq.~(\ref{rhfm}). Note that $\cos \theta \lesssim 0.1$ means that
the main origin of radiation is near the magnetic equatorial
plane, in accord with our previous discussion on the plasma
distribution. Second, our spectral power Eq.~(\ref{rfm}) displays
a spectral index $\alpha  \simeq - 1.33$ in reasonable agreement
with the observed typical spectral index. Moreover, the radial
distance $r$ cannot exceed the light cylinder radius $R_L$. So
that in general we have that $r \leq r_{break}$. It is natural to
identify $\nu_{break}$, the frequency corresponding to $r_{break}$
according to Eq.~(\ref{rfm}), with the frequency where the
observed radio spectrum displays a break. Usually it is found that
$\nu_{break} \simeq 1~ GHz$~\cite{graham-smith:2003}. Using the
typical pulsar parameters, Eq.~(\ref{typical}), we find $r_{break}
\simeq R_L/3$,  quite a reasonable result. To estimate the radio
luminosity:
\begin{equation}
\label{radio-lum}
L_{Radio} = \int_{\nu_{break}} F(\nu) \; d\nu \; \; ,
\end{equation}
where the integration extends up to a frequency $\nu \gg
\nu_{break}$, we  assume as typical electron number density $n_e
\simeq 10^{11} cm^{-3}$, and $v_\varphi \simeq 0.5$. In this way
we obtain:
\begin{equation}
\label{lum-num}
L_{Radio}  \; \simeq \; 1.3 \; 10^{27} \; erg/s \;
 \int_{1} \; \nu_{GHz}^{-4/3} \; d \nu_{GHz}
 \; ,
\end{equation}
or
\begin{equation}
\label{radio-lum-num}
 L_{Radio} \;  \simeq \;
3.9 \; 10^{27} \; erg/s \;  \; .
\end{equation}
The spin-down power  is given by:
\begin{equation}
\label{ener-rot-dot}
 - \; \dot{E}_{R} \; =  \; 4 \;
 \pi^2 \; I \; \frac{\dot{P}}{P^3} \; \simeq \; 3.95 \; 10^{32} \; erg/s \;
 ,
\end{equation}
so that we get:
\begin{equation}
\label{ratio-radio}
\frac{L_{Radio}}{ | \dot{E}_{R}|} \; \simeq \; \; 10^{-5} \; ,
\end{equation}
which is, indeed, the correct order of magnitude for typical
observed radio luminosities~\cite{manchester:1977,michel:1991}. \\
Essentially the same mechanism accounts for the high energy
emission. Indeed, according to our previous discussion, in the
region closer to the surface we expect that electrons will undergo
ultra relativistic motion with Lorentz factor $\gamma \gg 1$. In
this case the emitted radiation can be though of as a coherent
composition of contributions coming from the components of
acceleration parallel and perpendicular to the velocity. It turns
out, however, that the radiation is mainly due to the
perpendicular component. In the case of motion in a magnetic field
the radiation spectrum will be mainly at the
frequency~\cite{schwinger:1949}(see also
Ref.~\cite{wallace:1977}):
\begin{equation}
\label{omega_m}
\omega_m \; \simeq \; \gamma^2 \; \frac{e  B}{ m_e} \; .
\end{equation}
So that, according to Eq.~(\ref{mag-polar}) we have:
\begin{equation}
\label{omega_m-r}
\omega_m(r) \; \simeq \; \gamma^2 \; \frac{e  B_0}{ m_e} \;
 \left ( \frac{ R}{r} \right )^3 \; ,
\end{equation}
or
\begin{equation}
\label{r-omega_m}
 r  \; \simeq \;  R \; \gamma^{2/3} \; \left ( \frac{e  B_0}{ m_e}  \right )^{1/3}
 \frac{1}{\omega_m^{1/3}}  \; .
\end{equation}
Proceeding as before and using:
\begin{equation}
\label{dr-domega_m}
 dr \; \simeq \; \frac{R}{3} \; \gamma^{2/3} \;
 \left ( \frac{e  B_0}{  m_e}  \right )^{1/3}
 \frac{1}{\omega_m^{4/3}} \; d\omega_m\; ,
\end{equation}
we get the high energy spectral power:
\begin{equation}
\label{HE-spect-pow}
 F_{HE}(\omega) \simeq  \frac{2 \pi^2}{3} \;  n_e \; e B_0  \; \gamma^{2/3} \;
 R^4  \;  \frac{\dot{P}}{P}
 \left ( \frac{e  B_0}{  m_e}  \right )^{1/3} \frac{1}{\omega^{4/3}} \; .
\end{equation}
The high energy luminosity is:
\begin{equation}
\label{HE-lum}
L_{HE} \; = \; \int_{\omega_{HE}} F_{HE}(\omega) \; d\omega \; \;
,
\end{equation}
where again the integration extends up to $\omega \gg
\omega_{HE}$, and $\omega_{HE}$ is the high energy break frequency
emitted at  radial distance $r_{HE}$. It is reasonable to assume
that $r_{HE} \simeq 0.1 \, r_{break}$. We further assume
$\omega_{HE} \simeq 1 MeV$, which leads to the estimate $\gamma
\simeq 10^5$. Finally, using $n_e \simeq 10^{13} cm^{-3}$, we find
for the high energy luminosity:
\begin{equation}
\label{HE-lum-num}
 L_{HE} \;  \simeq \;
4.7 \; 10^{29} \; erg/s \;  \; ,
\end{equation}
which in turns leads to:
\begin{equation}
\label{ratio-HE}
\frac{L_{HE}}{ | \dot{E}_{R}|} \; \simeq \;  10^{-3} \; .
\end{equation}
We see that also our high energy luminosity Eq.~(\ref{ratio-HE})
compares rather well with observations. 

In summary, we have discussed a fair general and simple mechanism
for radio and high energy emission. Our results are based on the
induced azimuthal electric field which accounts for plasma
distribution in the region surrounding the pulsar, as well as for
the radio and high energy luminosities. We have also discuss the
formation of jet collinear with the magnetic axis, the pulsar wind
and the possible origin of the pulsar proper motion velocities.


\begin{thebibliography}{99}
%
\bibitem{hewish:1968}
A.~Hewish, S.~G.~Bell, J.~D.~H.~Pilkington, P.~F.~Scott, and
R.~A.~Collins, Nature\ {\bf 217}, 709 (1968).
%
%
\bibitem{baade:1934}
W.~Baade and F.~Zwicky,  Proc. \ Nat. \ Acad. \ Sci. \ {\bf 20},
254 (1934); Phys. Rev. {\bf 45}, 138 (1934); Phys. Rev. {\bf 46},
76 (1934).
%
%
\bibitem{manchester:1977}
R.~N.~Manchester and J.~H.~Taylor,
\newblock {\em Pulsars},
(W.~H.~Freeman and Company, San Francisco, 1977).
%
%
\bibitem{pacini:1968}
F.~Pacini,  Nature\ {\bf 219}, 145 (1968).
%
\bibitem{gold:1968}
T.~Gold,  Nature\ {\bf 218}, 731 (1968).
%
%
\bibitem{michel:1982}
See, for instance, F.~C.~Michel,  Rev. \ Mod. \ Phys. \ {\bf 54},
1 (1982); F.~C.~Michel, {\em The State of Pulsar Theory },
astro-ph/0308347.
%
%
\bibitem{michel:1991}
F.~C.~Michel,
\newblock {\em Theory of Neutron Star Magnetospheres},
(The University of Chicago Press, Chicago, 1991).
%
\bibitem{meszaros:1992}
P.~M\'esz\'aros,
\newblock {\em High-Energy Radiation from Magnetized Neutron Stars},
(The University of Chicago Press, Chicago, 1992).
%
%
\bibitem{goldreich:1969}
P.~Goldreich and W.~H.~Julian,  Astrophys.\ J.\  {\bf 157}, 869
(1969).
%
%
\bibitem{sturrock:1971}
P.~A.~Sturrock,  Astrophys.\ J.\  {\bf 164}, 529 (1971).
%
%
\bibitem{cea:2003}
P.~Cea, {\em P-Stars}, astro-ph/0301578.
%
%
\bibitem{cea:2004a}
P.~Cea, {\em {\it RXJ1856.5-3754} and {\it RXJ0720.4-3125} are
P-Stars }, astro-ph/0401339, to appear in JCAP.
%
%
\bibitem{cea:2004b}
P.~Cea, {\em Magnetic Fields and Glitches in P-Stars}, in
preparation.
%
%
%
\bibitem{ATNF}
http://www.atnf.csiro.au/research/pulsar/psrcat.
%
%
\bibitem{wallace:1977}
See, for instance:
%
W.~H.~Wallace,
\newblock {\em Radiation Processes in Astrophysics}
(MIT Press, Cambridge, 1977);
%
V.~L.~Ginzburg,
\newblock {\em Theoretical Physics and Astrophysics}
(Pergamon, Oxford, 1979).
%
\bibitem{pavlov:2001}
G.~G.~Pavlov, O.~Y.~Kargaltsev, D.~Sanwal, and G.~P.~Garmire,
Astrophys.\ J.\ {\bf 544}, L189 (2001).
%
\bibitem{helfand:2001}
D.~J.~Helfand, E.~V.~Gotthelf, and J.~P.~Halpern, Astrophys.\ J.\
{\bf 556}, 380 (2001).
%
\bibitem{hester:2002}
J.~J.~Hester, K.~Mori, D.~Burrows, J.~S.~Gallagher, J.~R.~Graham,
M.~Halverson, A.~Kader, F.~C.~Michel, and P.~Scowen, Astrophys.\
J.\ {\bf 577}, L49 (2002).
%
\bibitem{gaensler:2002}
B.~M.~Gaensler, J.~Arons, V.~M.~Kaspi, M.~J.~Pivovaroff, N.~Kawai,
and K.~Tamura, Astrophys.\ J.\ {\bf 569}, 878 (2002).
%
\bibitem{pavlov:2003}
G.~G.~Pavlov, M.~A.~Teter, O.~Kargaltsev, and D.~Sanwal,
Astrophys.\ J.\ {\bf 591}, 1157 (2003).
%
%
\bibitem{kijak:1997}
J.~Kijak and J.~Gil, MNRAS \ {\bf 288}, 631 (1997).
%
%
\bibitem{graham-smith:2003}
F.~Graham-Smith, Rep.\ Prog.\ Phys.\ {\bf 66}, 173 (2003).
%
%
\bibitem{schwinger:1949}
J.~Schwinger Phys. \ Rev. \ {\bf 75}, 1912 (1949).
%
%
%
\end{thebibliography}
\end{document}